\documentclass[
    ,final            
  ]
  {aipproc}

\layoutstyle{6x9}

\begin{document}

\title{In-Medium Vector Mesons, Dileptons and Chiral Restoration}


\author{Ralf Rapp}{
  address={Cyclotron Institute and Department of Physics \& Astronomy, 
Texas A\&M University, College Station, TX 77843-3366, USA}
}

\begin{abstract}
Medium modifications of the electromagnetic spectral function in
hadronic and quark-gluon matter are reviewed. A strong broadening of the 
$\rho$ meson, which dominates the spectral function in the low-mass 
regime, is quantitatively consistent with dilepton excess spectra 
measured in photoproduction off cold nuclei (CLAS/JLab) and in 
fixed-target ultrarelativistic heavy-ion collisions (NA45,NA60/CERN-SPS). 
The large excess observed by PHENIX at RHIC remains unexplained to date, 
but is most likely not due to emission from the Quark-Gluon Plasma. 
Connections to thermal lattice QCD promise progress in the search for 
chiral symmetry restoration.  
\end{abstract}

\maketitle


\section{Introduction}
The vacuum of Quantum Chromodynamics (QCD) is very dense, 
filled with condensates made of composite quark and gluon 
configurations. These condensates are believed to be at the origin of 
two of the most prominent nonperturbative phenomena of the strong 
interaction, namely the generation of hadronic masses and the 
confinement of color charge. For example, the formation of the chiral 
quark condensate, $\langle 0|\bar q q|0\rangle \simeq -2$\,fm$^{-3}$
per light quark flavor ($u$, $d$ and maybe $s$), breaks the chiral
invariance of the QCD Lagrangian and generates constituent
light-quark masses of $m_{u,d}\simeq350$\,MeV, compared to the bare
quark masses of $m_{u,d}^0\simeq5$-10\,MeV in the Lagrangian.     
The origin of confinement is less well understood; an intuitive (and
theoretically viable~\cite{Bonati:2010tz}) picture is that of a dual 
superconductor, where the condensation of color-magnetic monopoles 
forces color-electric into quasi-one-dimensional configurations,
the color-electric flux tubes (strings)  generating the confining
potential between two static color charges.   

\begin{figure}
\begin{minipage}{0.5\linewidth}
\includegraphics[height=9cm]{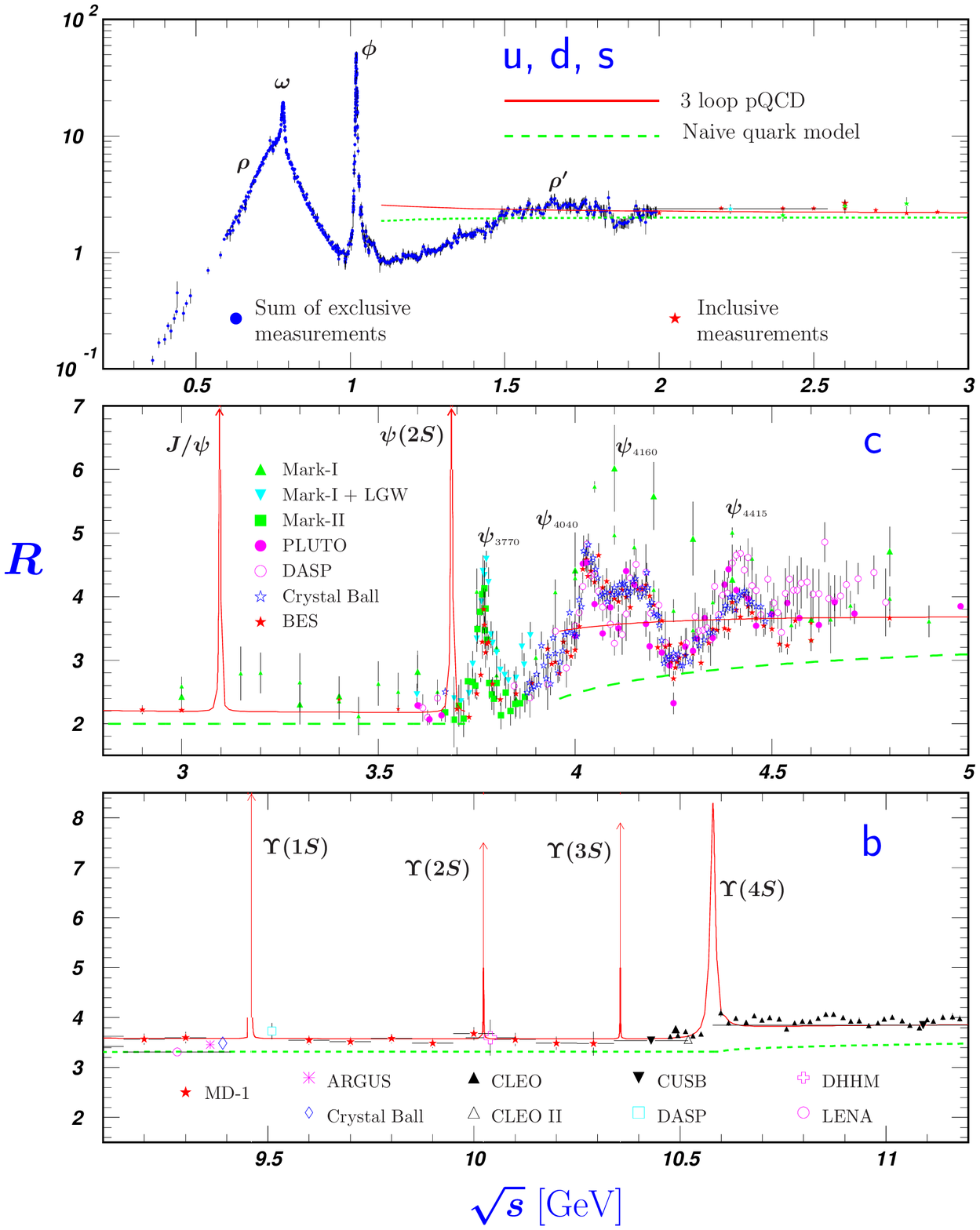}
\end{minipage}
\begin{minipage}{0.5\linewidth}
\includegraphics[width=6cm,angle=-90]{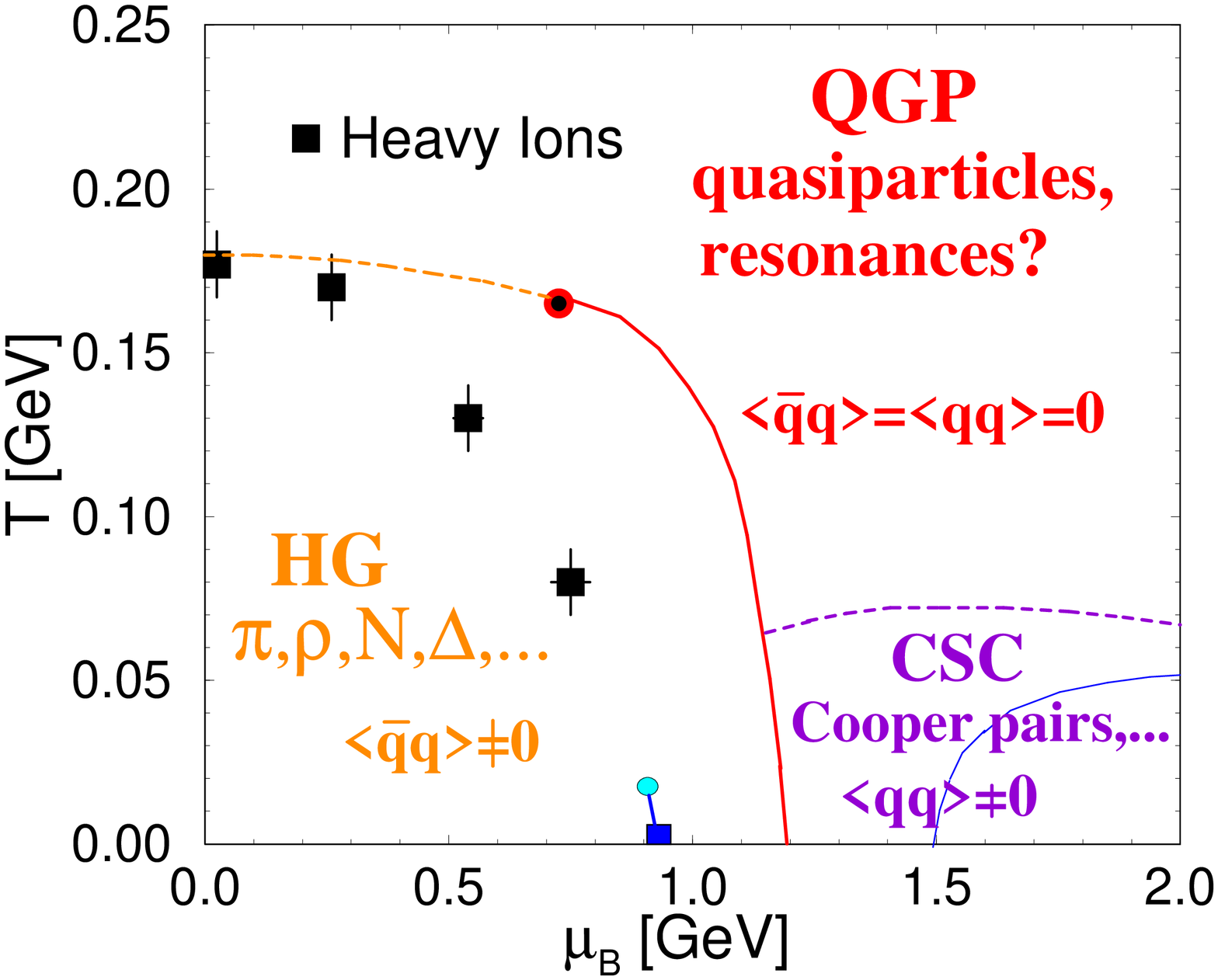}
\end{minipage}
\caption{Left panel: data compilation~\cite{Nakamura:2010zzi} of the 
cross-section ratio 
$R=\sigma(e^+e^-\to hadrons)/\sigma(e^+e^-\to \mu^+\mu^-)$
as a function of center-of-mass energy.
Right panel: Schematic diagram of the phase structure of QCD indicating
hadronic matter at low temperature ($T$) and baryon chemical potential
($\mu_B$), and transitions into a Quark-Gluon Plasma (QGP) at high $T$
and into Color-Super-Conducting (CSC) phases at high $\mu_B$ (small $T$).
For each of these 3 regions the lowest dimension quark condensates and
expected degrees of freedom are indicated. The idea is to infer these 
properties by using the vector spectral function shown in the left panel
as a probe.}
\label{fig_probe}
\end{figure}

Since condensates are not directly observable, one has to infer their 
presence and properties via their excitations. In the QCD vacuum these
are nothing but the hadron spectrum. A particularly instructive example 
is the vector ($J^{PC}=1^{--}$) spectral function which is accurately 
measured in electron-positron annihilation into hadrons, see left panel
in Fig.~\ref{fig_probe}\footnote{The vector and electromagnetic (EM) 
correlation functions essentially differ by a factor of the electric 
charge squared, $e^2$.}. 
The upper panel nicely illustrates a decomposition into a nonperturbative 
regime at masses below $\sim$1.5\,GeV and a perturbative regime above. In 
the former, the spectral strength is concentrated in the formation of the 
low-mass vector mesons ($\rho$, $\omega$ and $\phi$) with dynamical mass
while the latter is characterized by a continuum shape whose strength  
coincides with that of a perturbative $q\bar q$ final state (the 
subsequent formation of the multi-hadron final state appears to have 
little impact on the cross section)\footnote{In the vicinity of the 
charm-anticharm (middle panel) and bottom-antibottom thresholds (lower 
panel), heavy quarkonium bound states ($\Psi$ and $\Upsilon$ families) 
prevail, encoding information on the confining force.}. An analogous 
pattern is observed in the spacelike regime (negative $q^2\equiv -Q^2$), 
as probed, e.g., in electron scattering off the proton: JLab data for 
the $F_2^p$ structure function~\cite{Niculescu:2000tj} show a gradual 
transition from a nucleon-resonance dominated regime at low $Q^2$ to a 
structureless (perturbative) continuum for $Q^2\geq 3$\,GeV$^2$, in
agreement with the universal form measured in deep-inelastic scattering 
(DIS).    

A central objective in nuclear physics is the exploration of matter 
aggregates governed by the strong force, i.e., its phase diagram, 
schematically depicted in the right panel of Fig.~\ref{fig_probe}.
At high temperature (and baryon chemical potential) 
asymptotic freedom of QCD predicts the bulk medium to be a weakly 
interacting plasma of quarks and gluons (modulo colorsuperconducting 
quark pairing at a cold Fermi surface). Thus, hadronic spectral
functions should be close to their perturbative form, with no more
resonant correlations.    
For the vector spectral function shown in Fig.~\ref{fig_probe} this
implies that the resonance peaks have melted into a structureless 
continuum (signaling chiral restoration and deconfinement). A key 
question is how this transition is realized,
in particular when passing through phase changes of the medium.
This is the main idea of vector-meson spectroscopy of hot and 
dense matter. The vector channel is unique since it directly couples 
to (real and virtual) photons which can carry undistorted spectral 
information from the interior of strongly interacting systems in the 
laboratory (nuclei, heavy-ion collisions): the EM mean-free-path is 
much larger than the system size, $\lambda_{\rm em} \gg R_{\rm nucleus}$.

In the following sections we review calculations of $\rho$-meson spectral 
functions in hot and dense hadronic matter, discuss their applications to 
dilepton production experiments in photon-nucleus and ultrarelativistic 
heavy-ion collisions (URHICs), and elaborate connections to thermal 
lattice QCD. The final section contains conclusions.

\section{Vector Mesons in Vacuum and in Medium}
\label{sec_vmes}
Starting point for the description of vector-meson spectral
function in hadronic matter is a realistic model in vacuum. 
For the $\rho$-meson, on which we will focus here, this requires
to compute its coupling to 2-pion states to properly describe 
$P$-wave $\pi\pi$ phase shifts and the pion EM formfactor in the
timelike regime~\cite{Urban:1998eg,Harada:2003jx}.  

\begin{figure}[!t]
\begin{minipage}{0.5\linewidth}
  \includegraphics[width=0.83\textwidth,angle=-90]{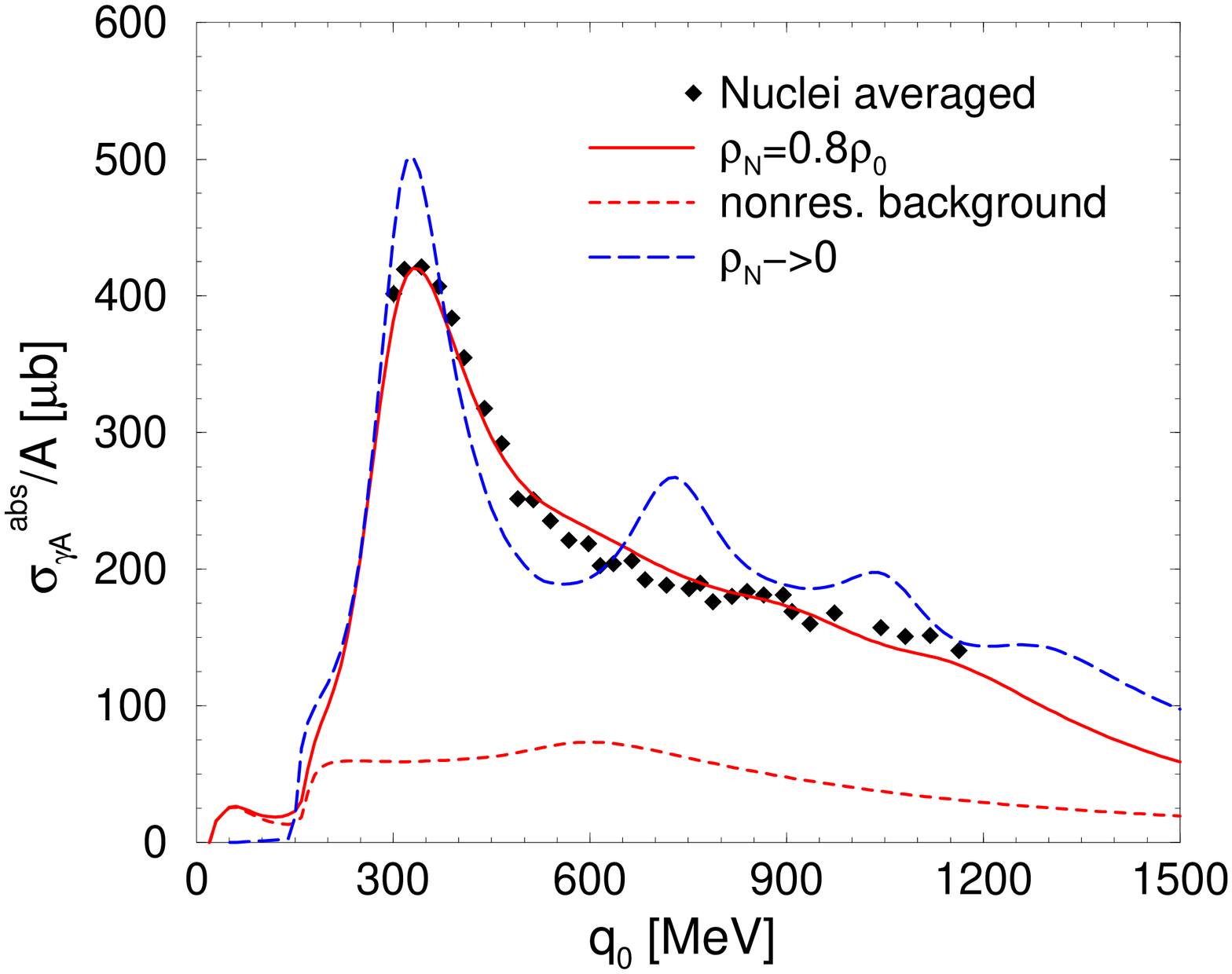}
\end{minipage}
\hspace{-0.0cm}
\begin{minipage}{0.5\linewidth}
\vspace{0.4cm}
  \includegraphics[width=0.97\textwidth]{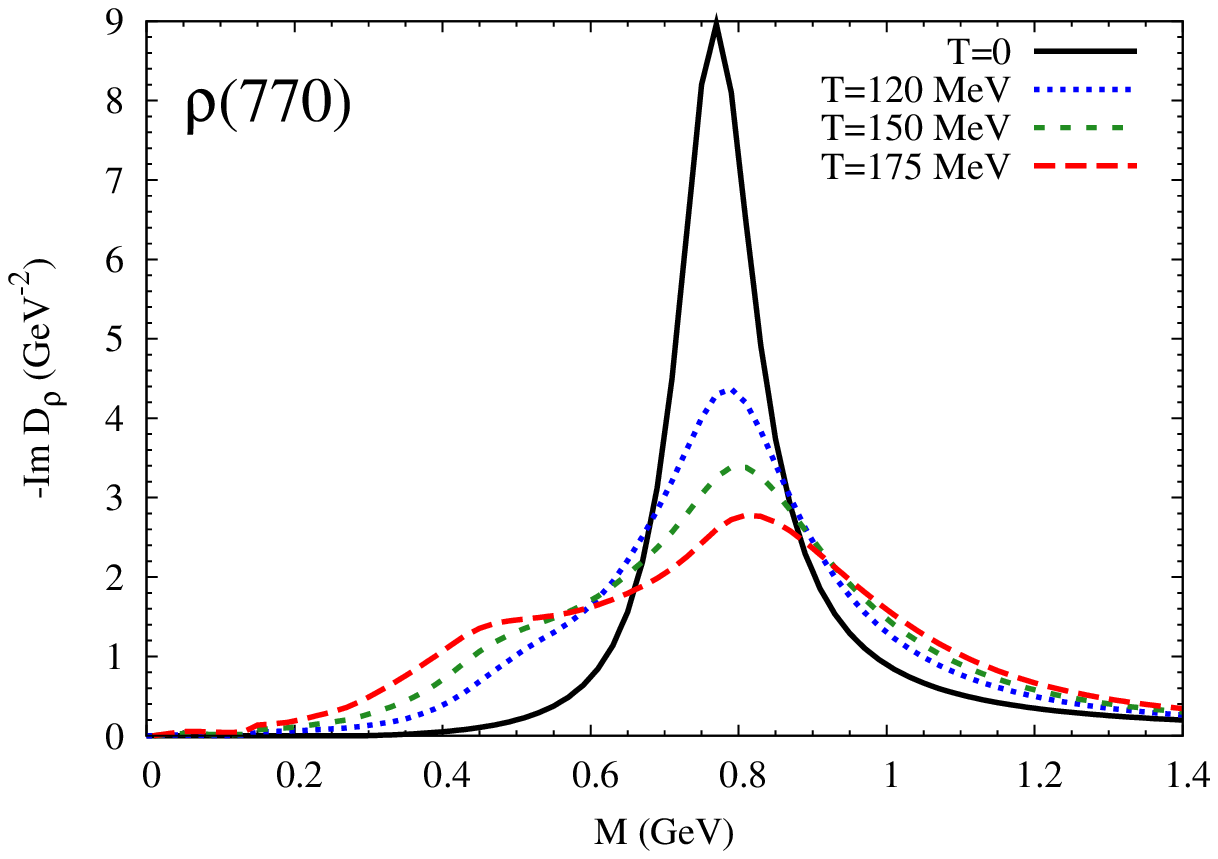}
\end{minipage}
  \caption{Left panel: constraints on the $\rho$-meson coupling to
nucleons by applying the light-like $\rho$ spectral function, 
Im\,$D_\rho(q_0=q,\varrho_N)$, to calculate total photoabsorption 
cross sections on the nucleon (low-density limit, long-dashed line) 
and on nuclei (solid line, for $\varrho_N=0.8\varrho_0$; short-dashed 
line: contribution from the in-medium pion cloud)~\cite{Urban:1998eg}. 
The data are averaged over nuclei from
Li to Pb (stat. error below 2\%). Right panel: in-medium $\rho$-meson
spectral function~\cite{Rapp:1999us} under conditions associated with 
heavy-ion collisions at the SPS; baryon-density effects are dominant, 
with $\rho_B$$\simeq$1.5(0.5)\,$\rho_0$ at $T$=175(120)\,MeV.}
\label{fig_Drho}
\end{figure}
Medium modifications of the $\rho$ propagator, $D_\rho=
[M^2-(m_\rho^0)^2-\Sigma_{\rho\pi\pi}-\Sigma_{\rho B,M}]^{-1}$,
are calculated via selfenergy insertions corresponding to a dressing 
of its pion cloud ($\Sigma_{\rho\pi\pi}$) and direct couplings to the
mesons ($\Sigma_{\rho M}$) and baryons ($\Sigma_{\rho B}$) in the heat 
bath (see Refs.~\cite{Rapp:2009yu,Leupold:2009kz} for recent reviews). 
To fix the coupling constants and formfactors of resonance 
excitations, empirical constraints from their decay branchings and 
scattering data are essential, e.g., using total photoabsorption cross 
sections on the nucleon and nuclei~\cite{Urban:1998eg}, cf.~left
panel of Fig.~\ref{fig_Drho}. 
The resulting $\rho$ spectral function, Im\,$D_\rho$, strongly broadens 
in the medium, approximately doubling (tripling) its free width in cold 
nuclear matter at half (full) saturation density ($\varrho_0=
0.16$\,fm$^{-3}$). Under conditions expected in heavy-ion collisions
at the CERN-SPS, the $\rho$ resonance ``melts'' (i.e., 
$\Gamma_\rho\to m_\rho$) when extrapolated into the putative phase 
transition region (see right panel of
Fig.~\ref{fig_Drho}), mostly driven by the baryonic component of the 
medium. This remains true at RHIC, despite the low 
net-baryon density at mid-rapidities, due to an appreciable density
of baryons {\em plus} antibaryons.

\section{Dilepton Phenomenology}
\label{sec_dilep}
Recent experiments have made major progress in extracting so-called 
``excess" dilepton spectra, which can be directly compared to 
theoretical calculations of the in-medium signal. This has been
achieved in both nuclear photoproduction at JLab~\cite{Wood:2008ee} 
(probing cold nuclear matter at $\varrho_N\approx0.5\varrho_0$) and in 
URHICs at the SPS~\cite{Arnaldi:2006jq,Adamova:2006nu,Arnaldi:2008er}
(probing hot and dense hadronic matter at 
$(\varrho_B,T)\approx(\varrho_0,150\,{\rm MeV})$).  
The main difference in the theoretical description of these
experiments lies in the excitation mechanism. In URHICs,
the large amount of incoming kinetic energy is quickly randomized 
producing a locally equilibrated (macroscopic) source, justifying the 
simplifying notion of thermal emission. In reactions with an elementary 
projectile (e.g., photon), the $\rho$ production process requires a 
microscopic calculation (e.g., $\gamma N\to\rho N$ amplitude), while
the subsequent in-medium propagation may be approximated assuming 
zero-temperature nuclear matter.

\subsection{Nuclear Photoproduction: Spectra and Absorption}
The CLAS experiment at JLab~\cite{Wood:2008ee} used photon 
Bremsstrahlung of energies $E_\gamma\simeq1$-3.5\,GeV to illuminate 
nuclear targets and measure
$e^+e^-$ spectra for $M_{ee}\simeq0.2$-1.2\,GeV.
After subtraction of the background and the narrow $\omega$ and $\phi$ 
peaks, the excess spectra were fitted with a schematic $\rho$-meson 
spectral function. For Fe-Ti targets, an average width of 
$\Gamma_\rho\simeq(220\pm15)$\,MeV was extracted, without significant 
mass shift.  

\begin{figure}
\begin{minipage}{0.5\linewidth}
  \includegraphics[width=1.0\textwidth]{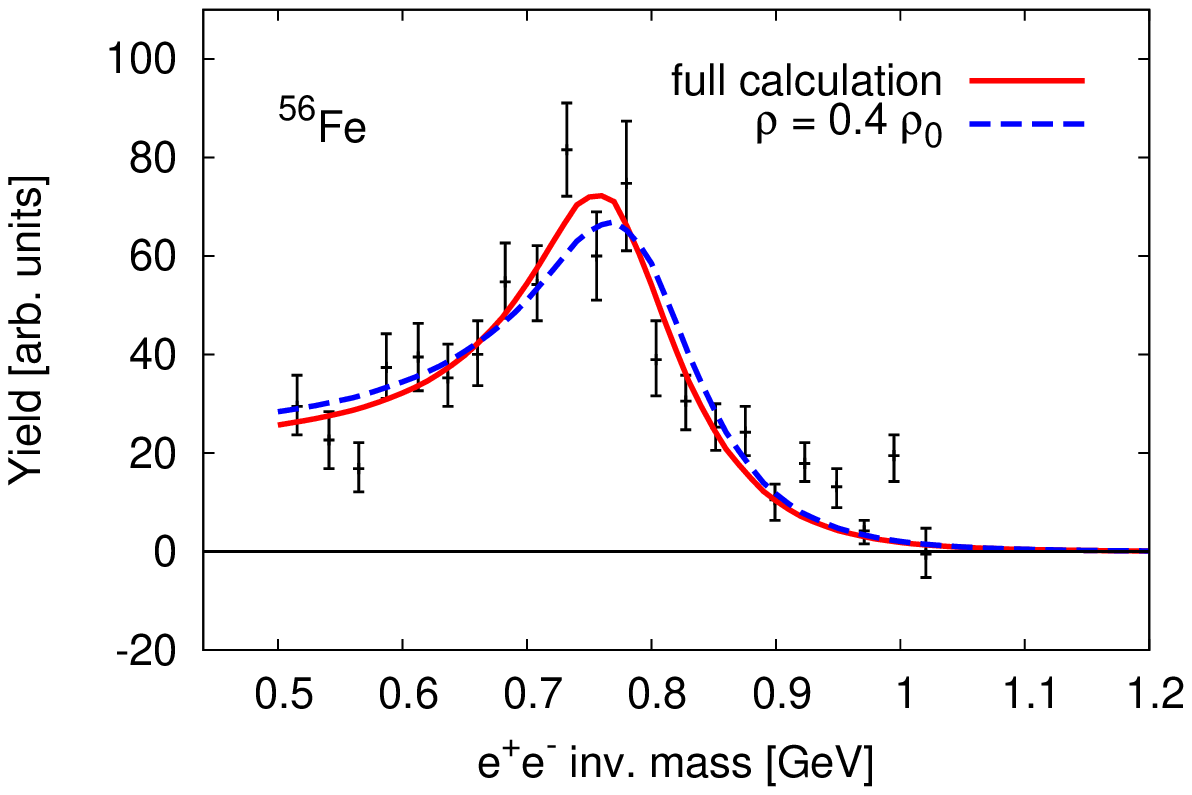}
\end{minipage}
\begin{minipage}{0.5\linewidth}
  \includegraphics[width=1.0\textwidth]{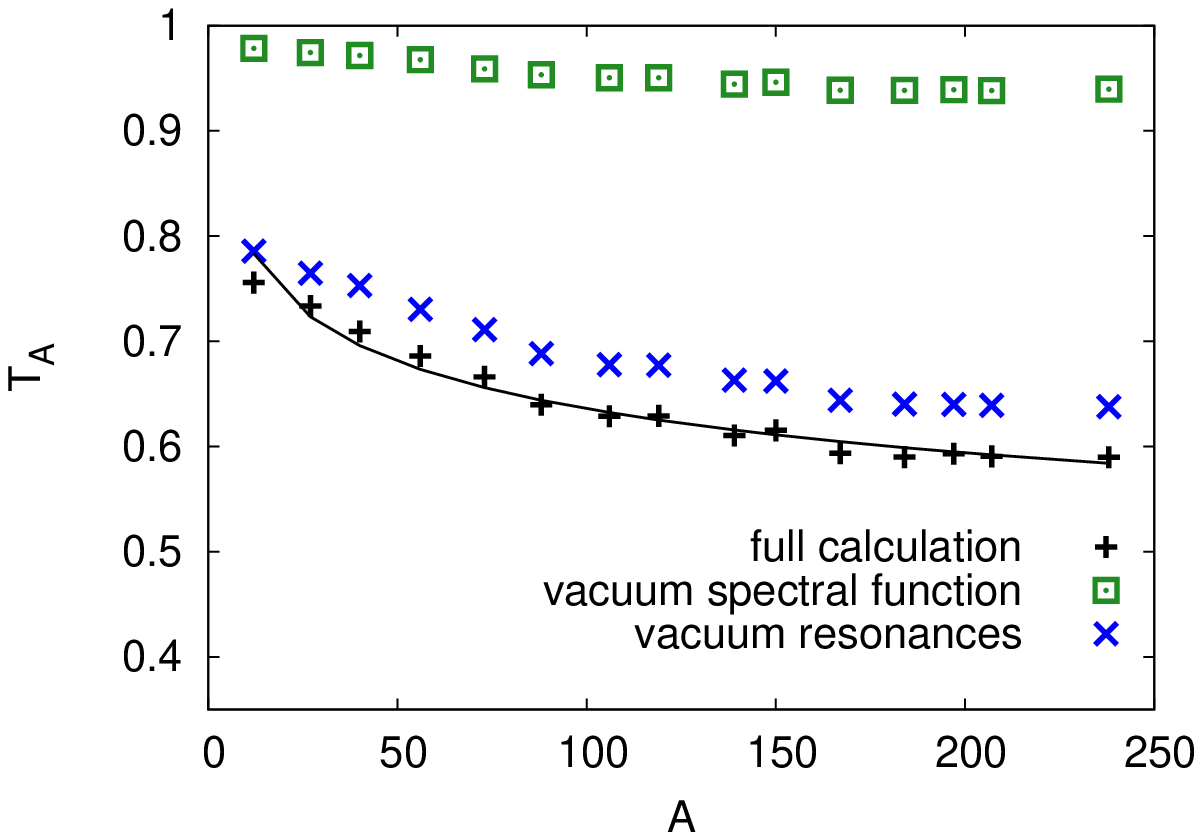}
\end{minipage}
\caption{Nuclear photoproduction of dileptons~\cite{Riek:2008ct} from 
$\rho$ decays employing the in-medium propagator of Fig.~\ref{fig_Drho} 
coupled with a realistic amplitude~\cite{Oh:2003aw} for the elementary 
$\gamma+N\to\gamma+\rho$ production process. Left panel: invariant-mass
spectra compared to CLAS data~\cite{Wood:2008ee}; right panel: predictions
for the nuclear transparency ratio as a function of nuclear mass number.}
\label{fig_clas}
\end{figure}

A calculation~\cite{Riek:2008ct} of the CLAS excess spectra using a 
realistic $\gamma N\to\rho N$ production amplitude~\cite{Oh:2003aw} 
coupled to in-medium $\rho$ propagation and decay~\cite{Rapp:1999us} 
is shown in Fig.~\ref{fig_clas}. The moderate broadening of the line
shape seen in the data is reasonably well reproduced without (re-) 
adjusting parameters. This may seem surprising given the large 
broadening of low-momentum $\rho$ mesons in nuclear matter as quoted 
above. However, the relatively large $\rho$ 3-momenta relative to the 
nucleus ($q\simeq1.5$\,GeV, inherited from the incoming photon energies) 
cause a reduction of medium effects since (a) a significant portion of 
$\rho$'s decay outside (or in the surface region) of the nucleus, leading 
to average densities at the decay point of $\bar\varrho\simeq0.4\varrho_0$ 
for $^{56}$Fe, and (b) the broadening of the spectral function decreases 
with increasing 3-momentum~\cite{Rapp:1999us}.
The significance of the medium effects can be quantified by a $\chi^2$
per data point of 1.08 when using the in-medium spectral function,
compared to 1.55 for the vacuum one. A complementary assessment of 
their magnitude can be obtained from the nuclear transparency ratio,
$T_A=\sigma_{\gamma A\to\rho X}/A\sigma_{\gamma N\to\rho X}$, which
reflects the absorptive width of the $\rho$ when traveling through the 
nucleus. The predictions shown in the right panel of Fig.~\ref{fig_clas}
show a nuclear absorption by 40\% for $A$$\simeq$200 nuclei, which is 
markedly less than what has been observed for $\phi$ and $\omega$ 
mesons~\cite{Ishikawa:2004id,Kotulla:2008xy}. The reason for this 
difference is not a smaller $\rho$ broadening in medium (in fact, it is
larger), but its large width in the vacuum, or rather the 
``anomalously'' small vacuum width of $\omega$ and $\phi$ (figuring 
into the denominator of $T_A$).

\subsection{Heavy-Ion Collisions: Thermal Radiation}
\begin{figure}[t]
\begin{minipage}{0.5\linewidth}
\hspace{0.1cm}
\includegraphics[width=0.93\textwidth]{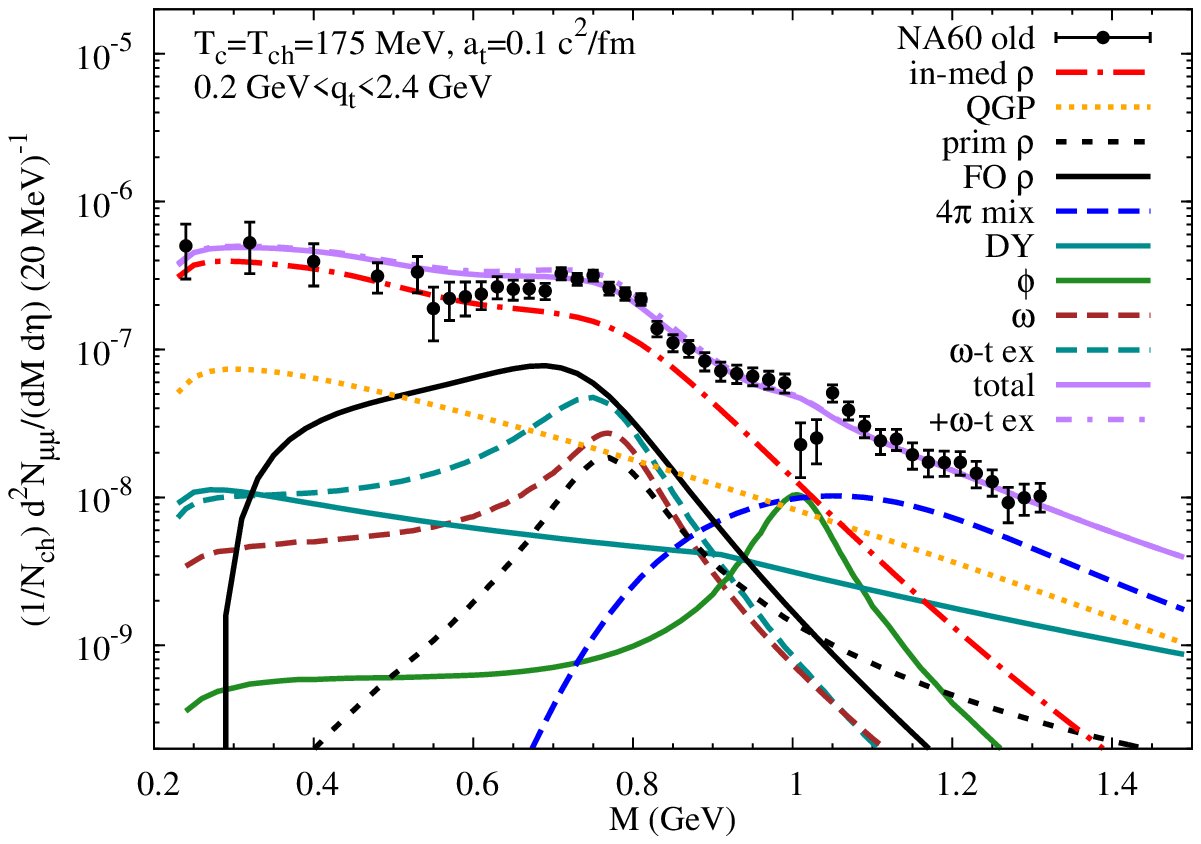}

\vspace{-0.55cm}

\includegraphics[width=0.78\textwidth,angle=-90]{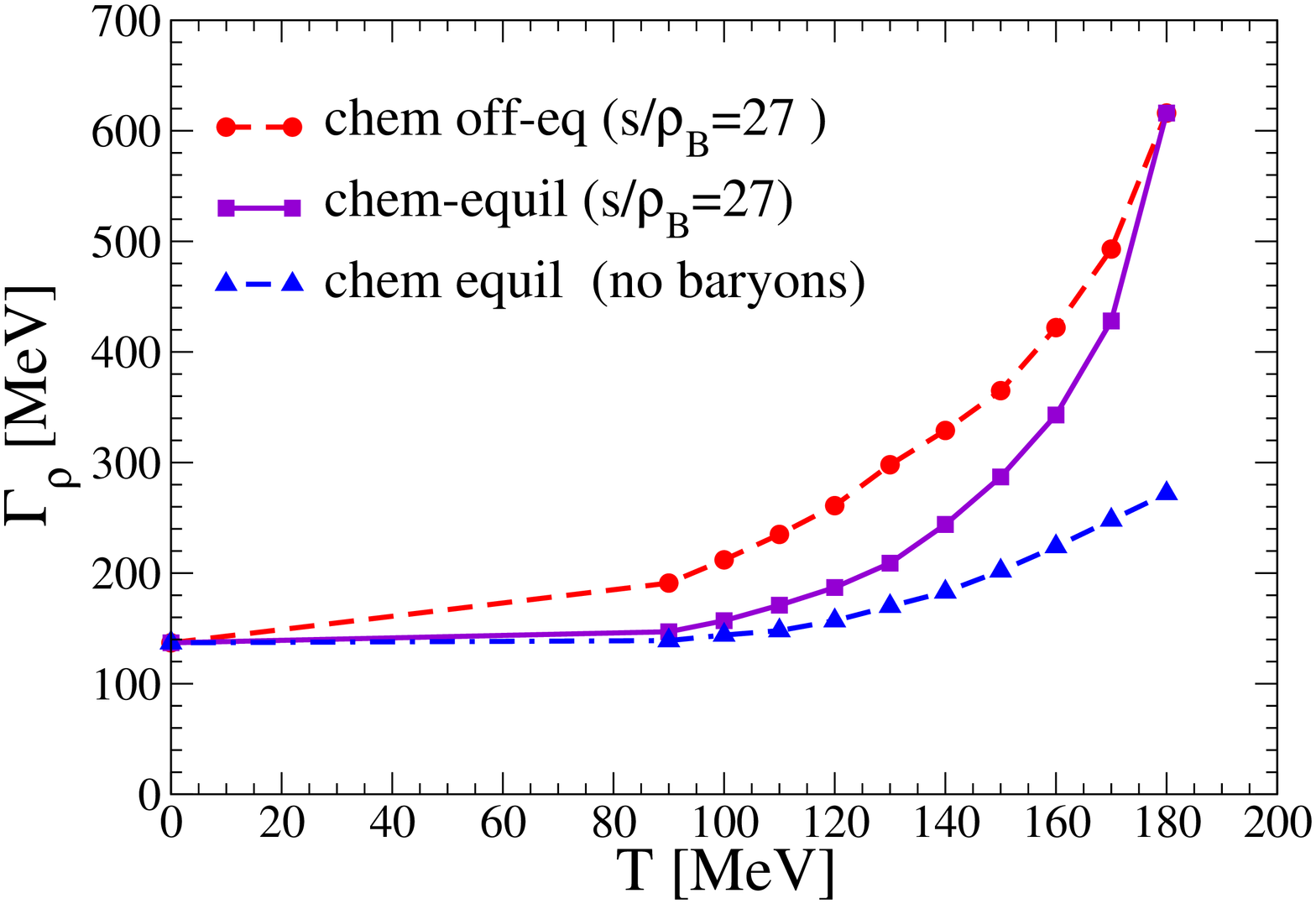}

\end{minipage}
\hspace{0.0cm}
\begin{minipage}{0.5\linewidth}
  \includegraphics[width=1.0\textwidth]{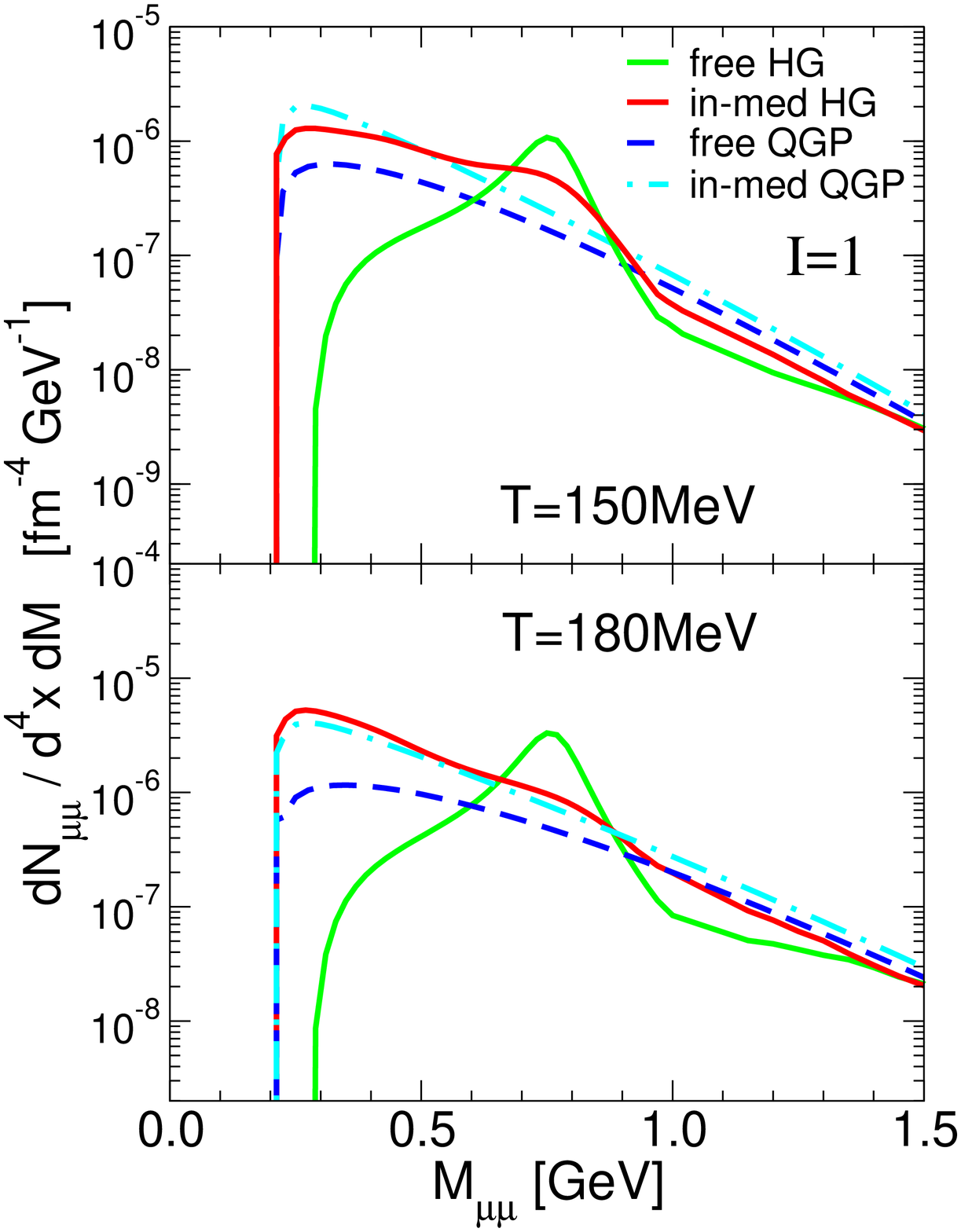}
\end{minipage}
\caption{Upper left: Comparison of theoretical calculations of dilepton 
invariant-mass spectra~\cite{vanHees:2007th} to fully acceptance-corrected
dimuon-excess spectra in In(158\,AGeV)-In collisions. Thermal sources
($\rho$ decays at low mass, multi-pion annihilation and QGP emission
at intermediate mass) dominate over non-thermal contributions.
Lower left: evolution of the $\rho$ width (circles) underlying the spectra
in the upper panel, compared to chemical equilibrium (squares) and when 
switching off interactions with baryons (triangles)~\cite{Rapp:2009yu}.
Right panel: 3-momentum integrated thermal dimuon emission rates for
isovector hadronic (solid red lines)~\cite{Rapp:1999us} and QGP emission 
(dash-dotted lines)~\cite{Braaten:1990wp} underlying the thermal spectra 
in the upper left, compared to free hadronic and $q\bar q$ rates.} 
\label{fig_na60}
\end{figure}

For the first time in heavy-ion collisions, the NA60 collaboration 
has recently achieved to extract a fully acceptance corrected 
{\em invariant}-mass spectrum of the observed dimuon-excess radiation 
in In-In collisions at the SPS (after subtraction of background and 
final-state hadron decays)~\cite{Arnaldi:2008er}, 
cf.~upper right panel in Fig.~\ref{fig_na60} (a further improved 
version of this spectrum can be found in Ref.~\cite{Damjanovic:2009zz}).
Theoretical calculations employing the in-medium $\rho$ spectral
function of Fig.~\ref{fig_Drho}, convoluted over an expanding fireball
model, compare well to the data~\cite{vanHees:2007th} (see also 
Refs.~\cite{Dusling:2007kh,Ruppert:2007cr,Bratkovskaya:2008bf}). The 
calculations are absolutely normalized and contain one main parameter, 
the fireball lifetime ($\tau_{\rm FB}\simeq6\pm1$\,fm/$c$), which 
governs the total yield in the spectrum. The shape of the spectrum is
determined by the predicted in-medium $\rho$ spectral function at low 
mass ($M\le1$\,GeV), and by continuum-like emission from the QGP and 
multi-pion fusion at intermediate mass ($M>1$\,GeV), with fixed 
relative strength. 
An important point here is that the Lorentz-invariant nature of the
$M$-spectra eliminates any distortion from the collective expansion
of the medium (pertinent blue shifts are known to strongly affect 
$p_t$-spectra of observed hadrons). This warrants a direct comparison 
of the experimental $M$-spectra to the theoretical input rates, 
\begin{equation}
\frac{dN}{d^4xdM} = -\frac{\alpha_{\rm em}^2}{\pi^3} \
\frac{L(M)}{M} \ \int \frac{d^3q}{q_0}f^B(q_0;T) \  \frac{1}{3} 
g_{\mu\nu}{\rm Im}~\Pi_{\rm em}^{\mu\nu} (M,q;\mu_B,T) \ ,
\end{equation}
which merely differ by the total four-volume of the expansion and the 
fact that the rates are for fixed temperature, while the spectra are
convoluted over it. Since higher masses are more sensitive to earlier
temperatures, one expects a larger average temperature to contribute 
at $M\simeq1.5$\,GeV than at $M\simeq0.3$\,GeV. Indeed, when evaluating 
the ratio of the spectrum for these masses, 
$\frac{dN}{dM}|_{0.3\,{\rm GeV}} / \frac{dN}{dM}|_{1.5\,{\rm GeV}}$, 
one obtains about $\sim$100, which is quite close to the ratio of the 
emission rate taken for $T=150$\,MeV at $M\simeq0.3$\,GeV and 
for $T=180$\,MeV at $M\simeq1.5$\,GeV.  
The lower left panel of
Fig.~\ref{fig_na60} shows the $\rho$ width as it evolves in the
medium at SPS (assumed to be in local thermal equilibrium). From the 
$M$-spectra one extracts an average value of 
$\bar\Gamma_\rho^{\rm med}\simeq 
\Gamma_\rho(\varrho_B$=$\varrho_0$,$T$=150{\rm MeV})\,$\simeq$\,350-400\,MeV, 
about 3 times the vacuum value. Given that roughly half of the 
radiation emanates from even higher $(\varrho_B,T)$ (as required, e.g., 
by the excess at $M>1$\,GeV), it is inevitable to conclude that the 
$\rho$ resonance indeed melts.

\begin{figure}[t]
\begin{minipage}{0.5\linewidth}
\includegraphics[width=0.94\textwidth]{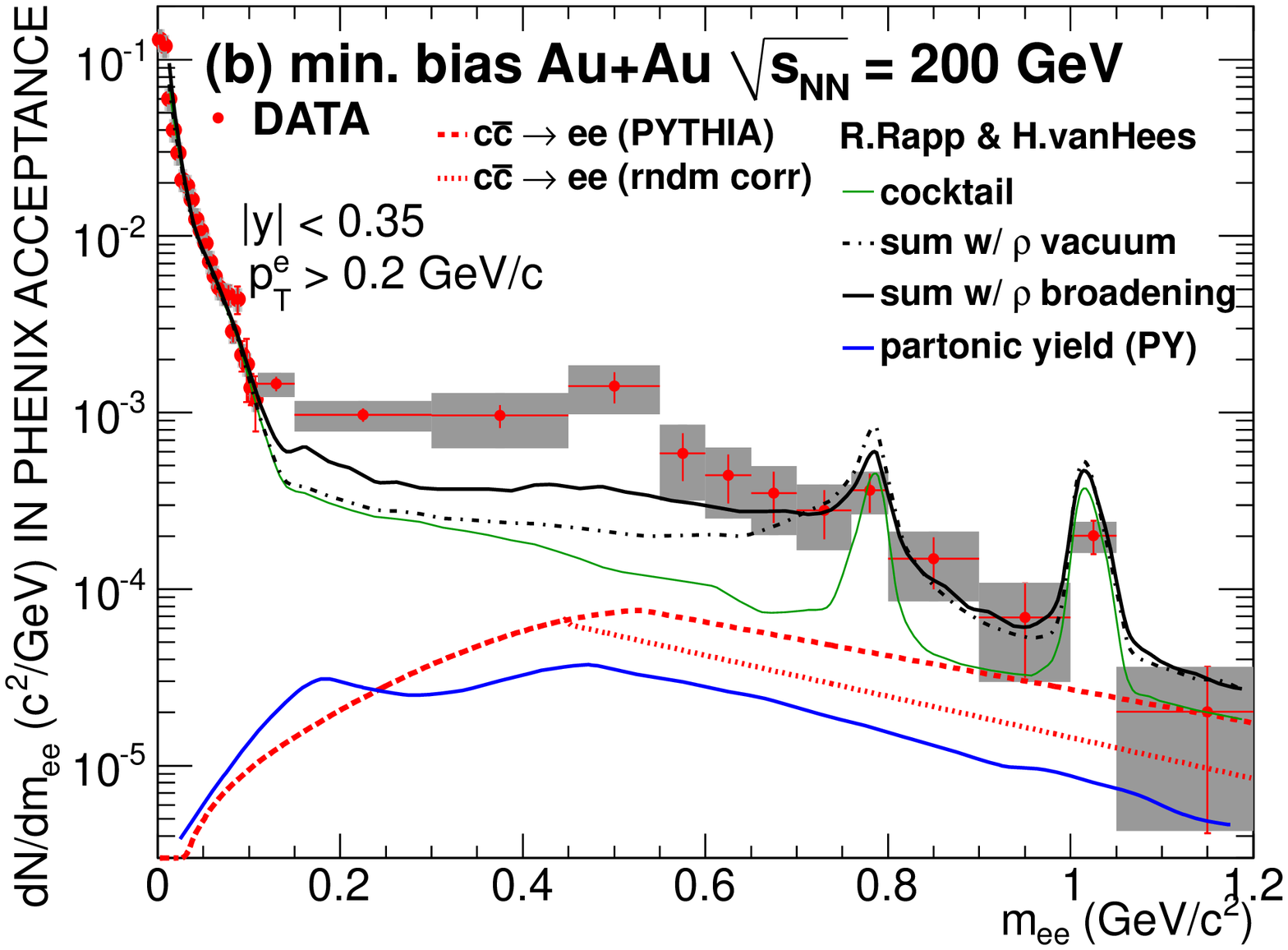}
\end{minipage}
\hspace{-0.4cm}
\begin{minipage}{0.5\linewidth}
\vspace{-0.35cm}
  \includegraphics[width=0.88\textwidth,angle=-90]{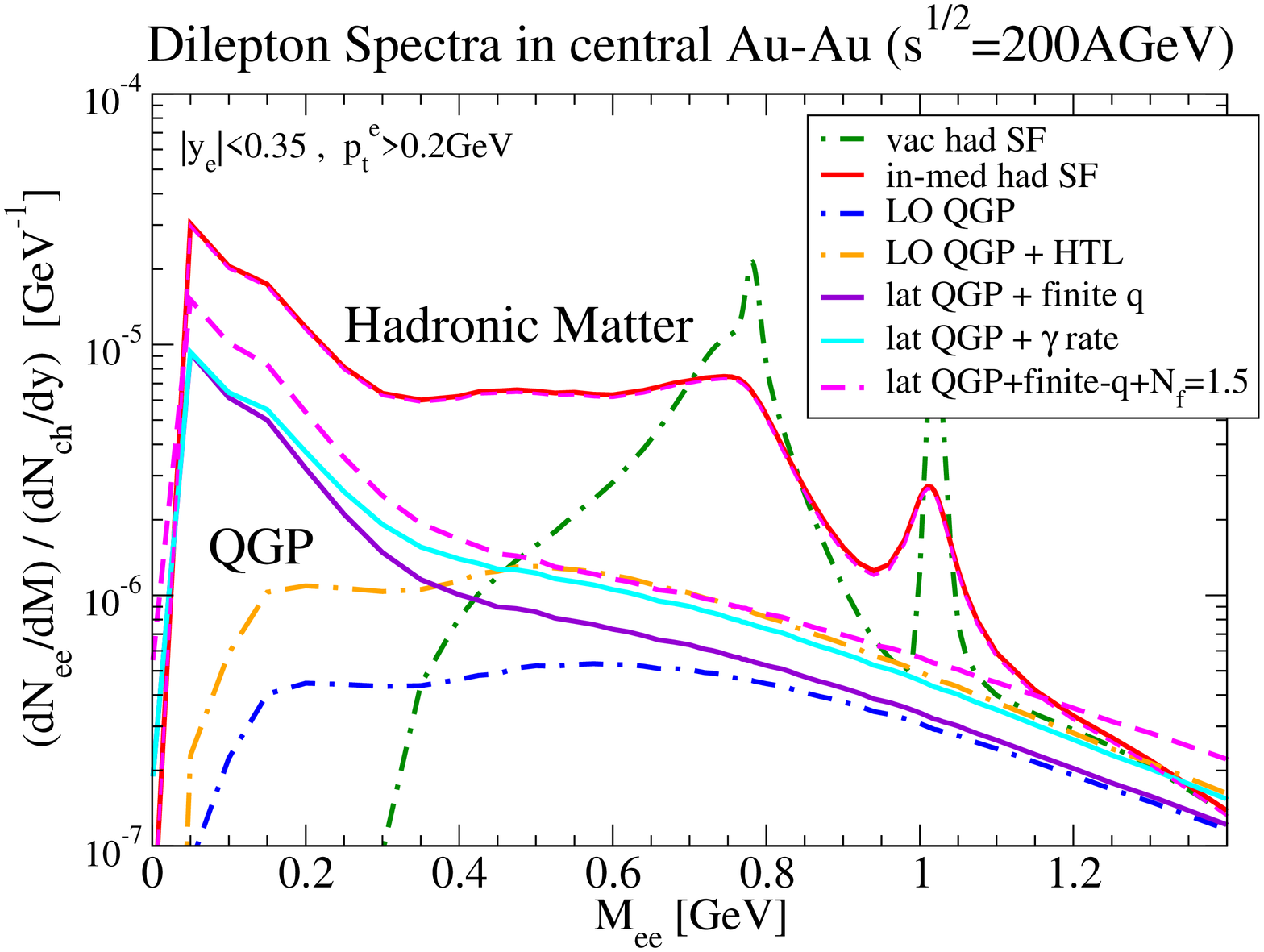}
\end{minipage}
\caption{Left panel: theoretical calculations of thermal dielectron spectra
in Au-Au collisions at RHIC using in-medium~\cite{Rapp:1999us} (upper
solid line) or vacuum (dash-dotted line) vector-meson spectral functions, 
added to QGP radiation and the cocktail of hadronic decays after thermal 
freezeout (including correlated charm decays)~\cite{Toia:2010}, and 
compared to PHENIX data~\cite{Adare:2009qk}. Right panel: studies of
QGP emission~\cite{Rapp:2010} including an improved photon limit with 
EM spectral functions fitted to recent lattice-QCD 
computations~\cite{Karsch:2010} (middle solid lines) and variations in 
the equation of state (dashed line).} 
\label{fig_phenix}
\end{figure}

Applying the same approach to di-electron spectra in Au-Au collisions
at a factor of $\sim$10 higher collision energies ($\sqrt{s}=200$\,AGeV 
at RHIC) leads to failure in comparison to recent PHENIX 
data~\cite{Adare:2009qk}, cf.~left panel of 
Fig.~\ref{fig_phenix}. At first one might think that the difference
arises from the factor of $\sim$2 higher initial temperatures 
at RHIC ($T_0\simeq 2T_c$) compared to SPS ($T_0\simeq T_c$). However, 
in the low-mass region, the thermal dilepton yield is not very sensitive
to the Boltzmann factor, but rather to the 4-volume of emission.
The latter is much smaller in the QGP than in the hadronic phase, and
this is the ultimate reason that QGP emission cannot compete
with thermal hadronic emission at masses around 0.3\,GeV. This is 
illustrated in the right panel of Fig.~\ref{fig_phenix}, where several
attempts have been made to augment the QGP contribution. None of these
reaches the size of the hadronic yield~\cite{Rapp:2010}. Thus, one is
led to conclude that the origin of the PHENIX enhancement must be a 
``cool'', long-lived hadronic source with little flow (as dictated by 
the small slope of the corresponding $q_t$ spectrum, 
$T_{\rm slope}\simeq 100$\,MeV). 
Together with further theoretical analysis, the upcoming PHENIX data for 
low-mass dileptons, which have been a priority of the recent
RHIC run-10, will hopefully shed light on this ``anomalous'' excess.  

Interesting results for dielectron spectra are also obtained at low 
energies, $E_{\rm lab}=$\,1-2\,AGeV~\cite{Agakishiev:2009yf}.
For light-ion projectiles (e.g., $^{12}$C), the dominant role seems
to be played by elementary processes, i.e., primordial $N$-$N$
Bremsstrahlung and final-state Dalitz decays of $\eta$, $\Delta(1232)$, 
etc. Heavier projectile-target configurations are hoped to reveal the
long-awaited results on vector-meson modifications in a low-$T$
high-$\varrho_B$ medium.

\section{Theoretical Interpretation}
\label{sec_theo}
\begin{figure}
  \includegraphics[width=6cm,angle=-90]{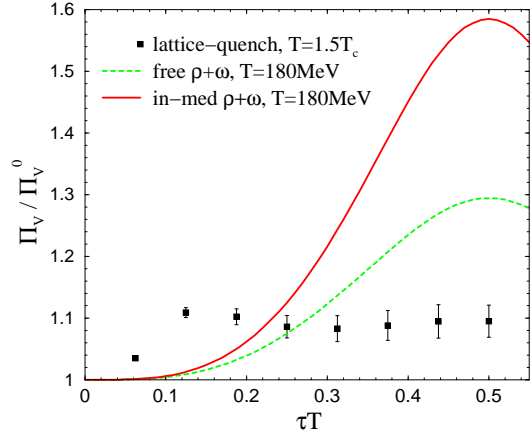}
\caption{Euclidean correlator ratios in the vector channel for the
hadronic many-body approach~\cite{Rapp:1999us} as computed in
Ref.~\cite{Rapp:2002pn}, compared to lattice QCD data of
Ref.~\cite{Karsch:2001uw}.}
\label{fig_Vcorr}
\end{figure}

Let us put the above findings into a broader perspective. Resonance
melting in the medium is a general phenomenon. It is visible in 
cold nuclei, where photo-absorption spectra exhibit the disappearance
of the second and third resonance region (recall left panel of 
Fig.~\ref{fig_Drho}; the $\Delta$(1232) width is ``protected'' by Pauli 
blocking in the $\pi N$ final state). Even the JLab data on the $F_2^N$ 
structure function on the deuteron indicate a ``premature'' approach to 
the general DIS scaling curve, i.e., at smaller $Q^2$ than on the 
proton~\cite{Niculescu:2000tj}. The same mechanism may very well be at
the origin of driving the timelike dilepton emission rate in hadronic
matter toward the one from hard-thermal-loop (HTL) resummed perturbative 
QCD~\cite{Braaten:1990wp} (recall the right panel of 
Fig.~\ref{fig_na60}). Note that
the perturbative result is automatically chirally symmetric. Further
constraints can be obtained from lattice QCD (lQCD) data for the euclidean 
correlator in the vector channel, which is related to the spectral 
function via
\begin{equation}
\Pi_V(\tau,q) = -\int\limits_0^\infty \frac{dq_0}{\pi} g_{\mu\nu} 
{\rm Im}\Pi_V^{\mu\nu}(q_0,q)
\frac{\cosh[q_0(\tau-1/2T)]}{\sinh[q_0/2T]} \ .
\end{equation}
Recent lQCD results, after removal of a so-called zero-mode 
contribution, find good agreement with the HTL result (except at 
small $q_0$ where HTL does not apply)~\cite{Karsch:2010}.   
Figure~\ref{fig_Vcorr} illustrates an early comparison~\cite{Rapp:2002pn}
of the euclidean correlator ratio for in-medium $\rho$ and $\omega$
spectral functions in the hadronic phase to quenched lQCD 
data~\cite{Karsch:2001uw} (the denominator, $\Pi_V^0$, corresponds to 
the non-interacting $q\bar q$ continuum for $u$ and $d$ flavors).
The agreement with the hadronic calculation improves for the recent,
zero-mode subtracted, lQCD data (the zero-mode contribution 
is not present in the hadronic calculation). 
Progress is also being made in evaluating axialvector spectral functions 
in medium~\cite{Cabrera:2009ep}, as reported at this 
meeting~\cite{Roca:2010}. A broadening found for the $a_1$ peak in 
nuclear matter supports a general tendency of ``chiral restoration 
through broadening''.

\section{Conclusions}
\label{sec_concl}
Hadronic many-body calculations of vector-meson spectral functions are
becoming a quantitative tool to perform spectral analysis of strongly 
interacting systems. Thus far, a strong broadening of the $\rho$-meson
consistently describes available dilepton and photon emission data in 
elementary and heavy-ion reactions. The only exception at current are 
the PHENIX low-mass dileptons, where a large QGP contribution is 
unlikely to resolve the discrepancy.  
In the absence of explicit measurements of the axialvector spectral
function, conclusions on chiral restoration have to be inferred 
indirectly. Degeneracy of the in-medium hadronic vector correlator with 
chirally symmetric perturbative and lattice spectral functions can, in 
principle, close this gap. 
Intriguing connections to the spacelike regime, accessible at JLab,
should be pursued further.



\begin{theacknowledgments}
I gratefully acknowledge my collaborators, specifically 
D.~Cabrera, C.~Gale, D.~Jido, T.S.H.~Lee, 
F.~Riek, L.~Roca, M.~Urban, H.~van Hees and J.~Wambach. 
This work has been supported by the U.S. National Science Foundation
under CAREER grant no.~PHY-0449489 and under grant no.~PHY-0969394,
and by the A.-v.-Humboldt Foundation. 
\end{theacknowledgments}



\bibliographystyle{aipproc}   


\begin{thebibliography}{9}

\bibitem{Bonati:2010tz}
  C.~Bonati {\it et al.}, 
  \emph{Phys.\ Rev.\  D} {\bf 81}, 085022 (2010).

\bibitem{Nakamura:2010zzi}
  K.~Nakamura {\it et al.}  [Particle Data Group],
  \emph{J.\ Phys.\ G} \textbf{37}, 075021 (2010).

\bibitem{Niculescu:2000tj}
  I.~Niculescu {\it et al.},
  \emph{Phys.\ Rev.\ Lett.}  {\bf 85}, 1182--1185 (2000).

\bibitem{Rapp:2009yu}
  R.~Rapp, J.~Wambach and H.~van Hees,
  in {\em Relativistic Heavy-Ion Physics}, edited by R.~Stock and 
  Landolt B\"ornstein (Springer), New Series {\bf I/23A}, 4-1 (2010);
  [{\tt arXiv:0901.3289[hep-ph]}].

\bibitem{Leupold:2009kz}
  S.~Leupold, V.~Metag and U.~Mosel,
  \emph{Int.\ J.\ Mod.\ Phys.\  E} {\bf 19}, 147--224 (2010).

\bibitem{Urban:1998eg}
  M.~Urban {\it et al.}, 
  \emph{Nucl.\ Phys.\  A} {\bf 641}, 433--560 (1998);
  R.~Rapp {\it et al.}, 
  \emph{Phys.\ Lett.\  B} {\bf 417}, 1--6 (1998).

\bibitem{Harada:2003jx}
  M.~Harada and K.~Yamawaki,
  \emph{Phys.\ Rept.}  {\bf 381}, 1-233 (2003).

\bibitem{Rapp:1999us}
R.~Rapp and J.~Wambach, 
\emph{Eur. Phys. J. A} \textbf{6}, 415--420 (1999).

\bibitem{Wood:2008ee}
  M.H.~Wood {\it et al.}  [CLAS Collaboration],
\emph{Phys.\ Rev.\  C} {\bf 78}, 015201 (2008).

\bibitem{Arnaldi:2006jq}
  R.~Arnaldi {\it et al.}  [NA60 Collaboration],
  \emph{Phys.\ Rev.\ Lett.}  {\bf 96}, 162302 (2006).

\bibitem{Adamova:2006nu}
D.~Adamova {\it et al.} [CERES/NA45 Collaboration],
\emph{Phys.\ Lett.\  B} {\bf 666},  425--429 (2008).

\bibitem{Arnaldi:2008er}
R.~Arnaldi {\it et al.}  [NA60 Collaboration],
  \emph{Eur.\ Phys.\ J.\  C} {\bf 59}, 607--623 (2009);
  \emph{ibid.} {\bf 61}, 711--720 (2009).

\bibitem{Riek:2008ct}
  F.~Riek {\it et al.}, 
  \emph{Phys.\ Lett.\  B} {\bf 677}, 116-120 (2009);
  \emph{Phys.\ Rev.\  C} {\bf 82}, 015202 (2010).

\bibitem{Oh:2003aw}
Y.~Oh and T.-S.H.~Lee,
\emph{Phys. Rev. C} {\bf 69}, 025201 (2004).

\bibitem{Ishikawa:2004id}
T.~Ishikawa {\it et al.},
\emph{Phys.\ Lett. B}  {\bf 608}, 215--222 (2005).
                                                         
\bibitem{Kotulla:2008xy}
M.~Kotulla {\it et al.}  [CBELSA/TAPS Collaboration],
\emph{Phys.\ Rev.\ Lett.}  {\bf 100}, 192302 (2008).

\bibitem{Damjanovic:2009zz}
  S.~Damjanovic, R.~Shahoyan and H.J.~Specht  [NA60 Collaboration],
  \emph{CERN Cour.}  {\bf 49N9}, 31 (2009).

\bibitem{vanHees:2007th}
H.~van Hees and R.~Rapp,
\emph{Nucl.\ Phys.\  A} \textbf{806}, 339--387 (2008). 

\bibitem{Dusling:2007kh}
  K.~Dusling and I.~Zahed,
  \emph{Phys.\ Rev.\  C} {\bf 80}, 014902 (2009).
                                              
\bibitem{Ruppert:2007cr}
  J.~Ruppert {\it et al.},
  \emph{Phys.\ Rev.\ Lett.}  {\bf 100}, 162301 (2008).

\bibitem{Bratkovskaya:2008bf}
  E.L.~Bratkovskaya, W.~Cassing and O.~Linnyk,
  \emph{Phys.\ Lett.\  B} {\bf 670}, 428--433 (2009). 

\bibitem{Braaten:1990wp}
  E.~Braaten, R.D.~Pisarski and T.C.~Yuan,
  \emph{Phys.\ Rev.\ Lett.}  {\bf 64}, 2242 (1990).

\bibitem{Adare:2009qk}
  A.~Adare {\it et al.}  [PHENIX Collaboration],
  \emph{Phys.\ Rev.\  C} {\bf 81}, 034911 (2010).

\bibitem{Toia:2010}
A.~Toia, private communication.

\bibitem{Rapp:2010}
R. Rapp, work in progress (2010).

\bibitem{Karsch:2010}
F. Karsch {\it et al.}, Proceedings of Lattice 2010, to be published. 

\bibitem{Agakishiev:2009yf}
  G.~Agakishiev {\it et al.}  [HADES Collaboration],
  \emph{Phys.\ Lett.\  B} {\bf 690}, 118--122 (2010).

\bibitem{Rapp:2002pn}
  R.~Rapp,
  \emph{Eur. Phys. J. A} \textbf{18}, 453--462 (2003).

\bibitem{Karsch:2001uw}
  F.~Karsch {\it et al.}, 
  \emph{Phys.\ Lett.\  B} {\bf 530}, 147--152 (2002).

\bibitem{Cabrera:2009ep}
  D.~Cabrera {\em et al.}, 
  \emph{Prog.\ Theor.\ Phys.}  {\bf 123}, 719--742 (2010).

\bibitem{Roca:2010}
L.~Roca, these proceedings.


\end{thebibliography}



\end{document}